\documentclass[letterpaper, 12pt, onehalfspacing, leqno]{article}

\usepackage{amssymb}
\usepackage{amsmath}
\usepackage{amsfonts}
\usepackage{amsthm}
\usepackage{graphicx}
\usepackage{enumerate}
\usepackage[margin=1.25in]{geometry}
\usepackage{setspace}
\usepackage[applemac]{inputenc}
\usepackage{dsfont}
\usepackage{natbib}
\usepackage{mathrsfs}
\usepackage{color}
\usepackage{comment}
\usepackage{latexsym}
\usepackage{cancel}
\usepackage{subcaption}
\usepackage{adjustbox}
\usepackage{booktabs}
\usepackage{tabularx}

\usepackage{geometry} 
\usepackage{xcolor, graphicx} 
\usepackage{tikz}
\usetikzlibrary{trees}

\usepackage{hyperref}
\usepackage{nameref}
\hypersetup{
  colorlinks   = true, 
  urlcolor     = blue, 
  linkcolor    = {violet!90!black}, 
  citecolor   = {gray}
}

\newtheorem{thm}{Theorem}
\newtheorem{prps}{Proposition}

\newtheorem{obs}{Observation}

\theoremstyle{definition}

\newtheorem{defn}{Definition}

\begin{document}
\title{The Focal Quantal Response Equilibrium}


\author{Matthew Kovach\footnote{Department of Economics, Purdue University.  E-mail: mlkovach@purdue.edu} \and Gerelt Tserenjigmid \footnote{Department of Economics, UC Santa Cruz.  E-mail: gtserenj@ucsc.edu}}

\date{Current Version: May 2026}

\maketitle

\begin{abstract} We propose a generalization of \emph{Quantal Response Equilibrium} (QRE) built on a simple premise: some actions are more \emph{focal} than others. In our model, which we call the \emph{Focal Quantal Response Equilibrium} (Focal QRE), each player plays a stochastic version of Nash equilibrium as in the QRE, but some strategies are focal and thus are chosen relatively more frequently than other strategies \emph{after accounting for expected utilities}.  The Focal QRE is able to systematically account for various forms of bounded rationality of players, especially regret-aversion, salience, or limited consideration. The Focal QRE is also useful for explaining the observed heterogeneity of bounded rationality of players across different games. We show that regret-based focal sets perform relatively well at predicting strategies that are chosen more frequently relative to their expected utilities. 
\end{abstract}

\noindent{{\it{Keywords: Focal Quantal Response Equilibrium; Stochastic Choice; Bounded Rationality; Regret Aversion; Preference Stability; Across Game Heterogeneity.}}
\bigskip


\section{Introduction}

In a Nash equilibrium of a normal-form game, each player follows their best response. However, there is an extensive literature documenting that play systematically deviates from Nash equilibrium (\cite{camerer2011behavioral}, \cite{goeree2016}). The \emph{Quantal Response Equilibrium} (QRE) of \cite{mckelvey1995quantal}  is a stochastic version of Nash equilibrium in which players do not choose the best response with probability one.  In the QRE, each player is a ``better responder'' rather than a best responder; agents play strategies with higher expected utility more frequently than those with lower utility. Therefore, the QRE circumvents the limitations of rational choice modeling of games by combining Nash equilibrium with \emph{random utility models}. More formally, player $i$ chooses strategy $s_{ij}$ if it maximizes the random expected utility of choosing $s_{ij}$ given the strategy profile of others, $p_{-i}$:
\[\max_{s_{ij}\in S_i}u_{i}(s_{ij}, p_{-i})+\epsilon_{ij}.\]

The QRE provides a statistical framework to analyze game-theoretic data from the field and laboratory and is one of the most successful and broadly used extensions of the Nash equilibrium (\cite{goeree2016}). The most important feature of the QRE is that it accounts for players' bounded rationality (and preference heterogeneity) by allowing them to follow random utility, while retaining the essence of Nash equilibrium.

However, there is abundant evidence from individual decision-making experiments and from experimental game theory that people fail to satisfy the standard assumptions of rational choice.  While some of these failures of rationality can be explained by the random utility framework (each agent maximizes a stochastic utility), there is convincing evidence that agents regularly make choices that are inconsistent with random utility models. For example, the \emph{attraction effect} (or decoy effect) and \emph{choice overload} are two well-known and robust violations of the random utility model. The attraction effect and choice overload always involve a violation of \emph{regularity}, a monotonicity property that must be satisfied by all random utility models, including multinomial logit and probit.\footnote{The attraction effect is exhibited when the introduction of an alternative $d_y$ to the menu $\{x, y\}$,  where $d_y$ is dominated by $y$ but not by $x$, increases the probability of choosing $y$. It was first documented experimentally by \cite{huber1982adding} and has been replicated by many studies in consumer choice (e.g.,  \cite{simonson1989choice}, \citealt{simonson1992choice}, \citealt{tversky1993context}, \citealt{ariely1995seeking}, \citealt{herne1998testing}, \citealt{doyle1999robustness}, \citealt{chernev2004extremeness}, and \citealt{sharpe2008using}). These effects are also demonstrated in the contexts of choice over risky alternatives (\citealt{herne1999effects}), choice over policy issues (\citealt{herne1997decoy}), choice over political candidates (\citealt{sue1995attraction}), among others. The attraction effect has also been observed in games, as documented by \cite*{colman2007asymmetric}. Choice overload involves an increasing tendency to either ``walk away'' from a choice or choose a default alternative as the number of available options increases and was first documented in \cite{iyengar2000choice}. Recently, \cite{de2024bounded} provide an excellent survey of the vibrant behavioral theory literature developed to explain violations of random utility models.}

The main contribution of this paper is a parsimonious generalization of the QRE called the \emph{Focal Quantal Response Equilibrium} (Focal QRE), which can account for such violations of the random utility model. Similar to the QRE, the Focal QRE keeps the equilibrium notion but relaxes the rationality assumption on the players' behavior. In a Focal QRE, it is as if each player divides her set of strategies into two groups: a focal group and a non-focal group. Strategies in the focal group are chosen excessively frequently when compared to other strategies \emph{after accounting for the expected utilities of these strategies}. Formally, in the Focal QRE,
player $i$ chooses strategy $s_{ij}$ if it maximizes the random expected utility of choosing $s_{ij}$ given the strategy profile of others, $p_{-i}$
\begin{equation}\max_{s_{ij}\in S_i}u_{i}(s_{ij}, p_{-i})+\delta_i\,\mathds{1}\{s_{ij}\in F_i\}+\epsilon_{ij},\end{equation}
where $\delta_i$ is the bias term for focal strategies and $F_i\subseteq S_i$ is the set of focal strategies. Separating the magnitude of focality bias, $\delta_i$, from focality, $F_i$, allows us to measure the extent to which focality influences behavior; it is natural to expect the impact of focality to depend on the game.\footnote{Similarly, \cite{kovach2017imbalanced} show that this separation is important to capture many empirical patterns in individual choice, especially the comparative statics of choice overload.} We show that the existence of the Focal QRE is ensured due to the properties of the QRE.

To facilitate application, we also introduce and discuss the logistic version of the Focal QRE, which we call the \emph{Focal Logit Equilibrium}.\footnote{The behavioral foundations of this choice behavior are provided in \cite{kovach2017imbalanced} in an abstract choice framework. They show that focal choice systematically accounts for the bounded rationality of individuals by nesting seemingly unrelated behavioral theories.} The logistic QRE is the most commonly applied specification because it is simple, widely applicable, makes clear predictions, and is often easy to estimate.  The Focal Logit Equilibrium also has similar advantages. We show in section \ref{identify} how focal strategies may be identified from frequency data under the assumption of Focal Logit Equilibrium.

To make predictions across different economic environments, economists often assume stability of preferences (or the distribution of preferences). Following this tradition, we assume stability of the distribution of preferences across different games, which is essential for the QRE to maintain testable implications across games. This is intuitively plausible in many settings, especially for strategically similar games with the same set of players. For instance, this is plausible when playing a symmetric or asymmetric matching pennies game in the lab. 
However, the QRE may struggle to accommodate changes in behavior across games when the error distribution is fixed. This may arise when behavior is close to Nash predictions in some games, but behavior is not close to Nash in other, strategically similar games. 

The Focal QRE can explain these changes in behavior across games with changing focality. Importantly, this allows us to maintain stability of preferences (i.e., a fixed error distribution and stable risk preferences across games) and therefore retain testable implications. In \autoref{hetero}, we illustrate how the Focal QRE can help explain across-game heterogeneity using data from the symmetric/asymmetric matching pennies games in \cite{goeree2001ten}, and the Attacker-Defender games in \cite{holt2024bilateral}.

The general Focal QRE framework does not require a single notion of focality. This flexibility is useful because it allows for the Focal QRE to apply in many different settings. Moreover, this generality enables the Focal QRE to facilitate the comparison of alternative theories of bounded rationality. However, it is helpful and instructive to introduce a specific theory of focality. Further, specifying a particular model of focality allows for sharper predictions and enhances the testability of the model. 

Accordingly, we introduce a theory of endogenous focality based on regret-aversion and show how to construct focal sets consistent with regret-averse behavior. This allows for the construction of focal sets directly from payoffs, which is useful for out-of-sample prediction and leads to sharper predictions. We find that regret-averse focality is mostly consistent with numerous ``puzzles'' in experimental games. We consider experimental data from \cite*{schotter1994laboratory} and the normal-form games from \cite{goeree2001ten}. Regret-averse focality is largely consistent with play in these games and accurately predicts when play will deviate from standard QRE predictions. 

 The remainder of the paper is organized as follows. After discussing related literature in \autoref{literature}, we formally define QRE in \autoref{section:QRE}. In \autoref{section:FQRE}, we introduce the Focal QRE and show equilibrium existence. In \autoref{hetero}, we show how the Focal QRE accounts for the heterogeneity of bounded rationality of players across different games. Finally, in \autoref{sec:regret}, we specialize the model and consider focal sets motivated by regret aversion. We conclude the paper by discussing different directions for endogenizing focal sets.

\subsection{Related Literature}\label{literature}

With mounting evidence that equilibrium theories often fail to explain or predict play in both the lab and the field, various models of boundedly rational play have been introduced. The QRE was among the first and has proven to be incredibly useful for explaining deviations from Nash equilibrium.\footnote{Other prominent models are ``Level $k$'' \citep{Stahl1994, Stahl1995, Nagel1995} and  ``Cognitive Hierarchy'' \citep{camerer2004cognitive}, both of which posit that players differ in their depth of reasoning.} The QRE has been widely applied to both normal form \citep{mckelvey1995quantal} and extensive form games \citep{mckelvey1998quantal}.

The QRE merges Nash equilibrium with taste shocks inspired by random utility models (RUM) \citep{falmagne1978representation, mcfadden1990stochastic}. The random utility framework has several interpretations. Most commonly, it is interpreted as a model of (i) preference heterogeneity (e.g., from a population), (ii) an individual with random preferences, or (iii) an individual reacting to information unobserved by the researcher.


Our paper fits into the literature on variations or extensions of the QRE. A recent example is the regular QRE of \cite{goeree2010regQRE}, which is a variation of the QRE where quantal response functions satisfy four regularity conditions, including monotonicity. In our model, monotonicity can be violated due to focality. 

Other variations and extensions include the (Subjective) Heterogeneous Quantal Response of \cite{Rogers2009}, which generalizes the logit specification of QRE to allow for players to differ in their ability and have incorrect beliefs about others' ability;\footnote{The authors also show that the `Cognitive Hierarchy'' model is a limit case of their model.} the Logit QRE with an endogenous $\lambda$ of \cite{friedman2020}; and the Symmetric QRE of \cite{friedman2022}, which is between the Logit QRE and the Regular QRE by introducing a symmetry axiom. 

Many of these variations focus on delivering models that are between the Logit QRE, which is quite restrictive, and the unrestricted QRE. Our paper, however, departs from these by directly relaxing the random utility requirement. Even when we assume logistic responses in the Focal QRE, yielding the Focal Logit Equilibrium, our model is more general than the Logit QRE and may not be a random utility model due to changes in focality across games.  Our paper, therefore, continues the tradition of relaxing the Logit QRE while restricting how play changes across games.

\section{Quantal Response Equilibrium}\label{section:QRE}

In this section, we closely follow section 2 of \cite{mckelvey1995quantal} to define QRE for normal-form games.

Consider a finite, $n$-person game in normal form: There is a set $N=\{1, 2, \ldots, n\}$ of players, and for each player $i\in N$ a strategy set $S_i=\{s_{i1}, \ldots, s_{iJ_i}\}$ consisting of $J_i$ pure strategies. Using a standard abuse of notation, $J_i$ also represents the set $\{1, \ldots, J_i\}$ of indices. For each $i\in N$ there is payoff function $u_i: S\to\mathds{R}$, where $S=\prod_{i\in N} S_i$. 

Let $\Delta_i$ be the set of probability measures on $S_i$. Elements of $\Delta_i$ are of the form $p_i:S_i\to[0, 1]$ where $\sum_{s_{ij}\in S_i} p_i(s_{ij})=1$. We write $\Delta=\prod_{i\in N}\Delta_i$ and let $J=\sum_{i\in N} J_i$. Abusing notation, we let $s_{ij}$ denote the strategy $p_i\in\Delta_i$ with $p_{ij}=1$. We also use the shorthand notation $p=(p_i, p_{-i})$. Hence, $(s_{ij}, p_{-i})$ represents the strategy where $i$ adopts the pure strategy $s_{ij}$, and all other players adopt their components of $p=(p_i, p_{-i})$. The payoff function is extended to have domain $\Delta$ by the rule $u_i(p)=\sum_{s\in S} p(s)u_i(s)$, where $p(s)=\prod_{i\in N}p_i(s_i)$. A vector $p=(p_1, \ldots, p_n)\in \Delta$ is a \emph{Nash equilibrium} (NE) if for all $i\in N$ and all $p'_i\in\Delta_i$, $u_i(p'_i, p_{-i})\le u_i(p)$.       

We write $X_i=\mathbb{R}^{J_i}$ to represent the space of possible payoffs for strategies that player $i$ might adopt, and $X=\prod_{i\in N} X_i$. We define the function $\bar{u}_i:\Delta\to X_i$ by 
\begin{equation}
\bar{u}_i(p)=\big(\bar{u}_{i1}(p), \ldots, \bar{u}_{iJ_i}(p)\big),\end{equation}
where 
\begin{equation}\bar{u}_{ij}(p)=u_i(s_{ij}, p_{-i}).\end{equation}

\cite{mckelvey1995quantal} define the QRE as a statistical version of NE where each player's utility for each action is subject to random error (i.e., random utility shock). Player $i$'s error vector, $\epsilon_i=(\epsilon_{i1}, \ldots, \epsilon_{i J_i})$, is distributed according to a joint distribution with density function $f_i(\epsilon_i)$. We call $f=(f_1, \ldots, f_n)$ \emph{admissible} if for each $i\in N$, the marginal distribution of $f_i$ exists for each $\epsilon_{ij}$ and $\mathbb{E}(\epsilon_{ij})=0$. 

The behavioral assumption of the QRE is that each player selects strategy $s_{ij}$, such that $\bar{u}_{ij}+\epsilon_{ij}\ge \bar{u}_{ik}+\epsilon_{ik}$ for all $k\in J_i$ as in random utility models. Notice that for any given $\bar{u}$ and $f$, the behavioral assumption implies a probability distribution over the observed actions of the players, induced by the probability distribution over the error vector, $\epsilon$. Let 
\[\sigma_{ij}(\bar{u}_i)=\int \mathds{1}\{\epsilon\big{|} \bar{u}_{ij}+\epsilon_{ij}\ge \bar{u}_{ik}+\epsilon_{ik}\text{ for each }k\in J_i\} f(\epsilon) d\epsilon,\] 
be the probability that player $i$ will select strategy $s_{ij}$ given $\bar{u}$.  We then define a \emph{Quantal Response Equilibrium} for any $f$ and $\Gamma=(N, S, u)$.

\begin{defn}[QRE] Let $\Gamma=(N, S, u)$ be a game in normal form, and let $f$ be admissible. A \emph{quantal response equilibrium} (QRE) is any $\pi\in\Delta$ such that for all $i\in N$ and $j\in J_i$,
\[\pi_{ij}=\sigma_{ij}(\bar{u}_{i}(\pi)).\] 

\end{defn}

\medskip
We call $\sigma_i:X_i\to \Delta^{J_i}$ the quantal response function of player $i$. The quantal response functions are non-empty and continuous. Hence, by Brouwer's fixed point theorem, the QRE exists. The utility of choosing $s_{ij}$, $\bar{u}_{ij}(p)=u_{i}(s_{ij}, p_{-i})$, is calculated using expected utility theory. However, it is immediate that the QRE can be defined for any utility theory as long as $\bar{u}_{ij}(p)$ is well-defined. However, for equilibrium existence, $\bar{u}_{ij}(p)$ must to be continuous in probabilities to use Brouwer's fixed point theorem. We now introduce our generalization of the QRE, called the \emph{Focal Quantal Response Equilibrium}. 

\section{Focal Quantal Response Equilibrium}\label{section:FQRE}

In the QRE, all strategies are treated equally after taking their expected payoffs into account. However, there is abundant evidence from the experimental literature on individual decision making that individuals treat choices differently independent of their payoffs or utilities. There is also robust evidence against the random utility model. Motivated by this literature, we assume that players act as if they have adopted the following procedure. Each player $i$ divides the set of strategies $S_i$ into two groups. One group is focal, and these focal strategies are chosen relatively more frequently than strategies in the non-focal group. Player $i$ might focus on a subset of $S_i$ for a multitude of reasons.  Throughout the paper we will focus on the following three behavioral interpretations.

\begin{itemize}
\item \textbf{Regret:} Some strategies are focal because they may lead to low ex-post regret.
\item \textbf{Salience:} Some strategies are salient because they are normatively appealing, easier to notice or evaluate, or involve payoffs that capture attention.\footnote{One prominent notion of salience is due to \cite{bordalo2013salience}.}
\item \textbf{Limited Consideration:} It is psychologically or physically too demanding for player $i$ to consider all strategies and their contingencies carefully. Therefore she focuses on a subset of $S_i$.  
\end{itemize}

To model the above procedure, we define a \emph{focus function}, which specifies the focal subset of $S_i$, and a \emph{distortion function}, which measures the bias toward strategies in the focal group.  The way players split $S_i$ and the magnitude of the bias toward the focal set may vary depending on the nature of the game.

Let $\mathcal{G}$ be the set of possible $n$-player normal-form games $\Gamma$ that players in $N$ can play. Let $\mathscr{S}_i$ be the collection of all possible sets $S_i$ of pure strategies that player $i$ can play games in $\mathcal{G}$. A mapping $F_i: \mathcal{G}\to \mathscr{S}_i$ is a \emph{focus function for player $i$} if $F_i(\Gamma)\subseteq S_i$ for all $\Gamma=(N, S, u)\in \mathcal{G}$. Therefore, $F_i(\Gamma)$ is the set of focal strategies for player $i$ in $\Gamma$. It is natural to have different focal sets for different players for asymmetric games since they have different roles. A \emph{distortion function} is a mapping $\delta_i\!:\!\mathcal{G}\to\mathds{R}_{+}$. In the Focal QRE, the utilities of strategies in $F_i(\Gamma)$ are biased by $\delta_i(\Gamma)$. We can now formalize the utility model consistent with the above procedure.

\begin{defn}[Focal Utility] For a given normal-form game $\Gamma=(N, S, u)$, the \textbf{focal expected utility} for $(F_i, \delta_i)_{i\in N}$ is defined by 
\begin{equation}
u^*_i(p)=\big(u^*_{i1}(p), \ldots, u^*_{iJ_i}(p)\big),\end{equation}
where 
\begin{equation}\label{focal:eq}u^*_{ij}(p)=u_i(s_{ij}, p_{-i})+\delta_i(\Gamma)\mathds{1}\{s_{ij}\in F_i(\Gamma)\}.\end{equation}

\end{defn}

In the context of individual decision making, \cite{kovach2017imbalanced} provide the behavioral foundations of the utility representation in \autoref{focal:eq}, as well as the version with logistic error. Moreover, we provide a formal connection between focal utility and theories of bounded rationality (salience theory, limited attention, etc.) and the nested logit.

An attractive feature of the QRE is that it retains the equilibrium notion but accounts for the bounded rationality of players by allowing them to have random utility. Similar to the QRE, we keep the equilibrium notion but we further relax the rationality assumptions. In particular, we allow for behavior that is inconsistent with random utility models (e.g., attraction effect) by introducing focality-adjusted utility.

As in the QRE, the probability that player $i$ will select strategy $s_{ij}$ given focal expected utility $u^*_i$ is \begin{equation}
\sigma_{ij}(u^*_i)=\int \mathds{1}\{\epsilon\big{|} u^*_{ij}+\epsilon_{ij}\ge u^*_{ik}+\epsilon_{ik}\text{ for each }k\in J_i\}f(\epsilon)\, d\epsilon.\end{equation}
Finally, we can define our generalization of the QRE for any admissible $f$, normal-form game $\Gamma=(N, S, u)$, and pair $(F_i, \delta_i)_{i\in N}$.

\begin{defn}[Focal QRE] Let $\Gamma=(N, S, u)$ be a normal-form game, and let $f$ be admissible. A \textbf{focal quantal response equilibrium} (Focal QRE) for a given $(F_i, \delta_i)_{i\in N}$ is any $\pi\in\Delta$ such that for all $i\in N$ and $j\in J_i$,
\[\pi_{ij}=\sigma_{ij}\big(u^*_{i}(\pi)\big).\] 
\end{defn}

Notice that the difference between the QRE and the Focal QRE is that the latter allows for focality to influence behavior in addition to utilities. Indeed, when $\delta_i(\Gamma)=0$ or $F_i(\Gamma)=S_i$, the Focal QRE reduces to the QRE. Similar to Theorem 1 of \cite{mckelvey1995quantal}, the Focal QRE exists in any normal-form game by Brouwer's fixed point theorem. The main reason that we are assured existence is that $F_i(\Gamma)$ and $\delta_i(\Gamma)$ are fixed before players consider strategic interactions. 

\begin{thm} For any $(F_i, \delta_i)_{i\in N}$, $\Gamma$, and admissible $f$, there exists a Focal QRE.  
\end{thm}

\subsection{Focal Logit Equilibrium}

Logit (or the Luce model) is the most well-known special case of RUM. Naturally, the most popular version of the QRE is the logistic QRE, as it inherits attractive properties of logit that are convenient in applications. The usefulness of the logit QRE is widely known in the literature (see \cite{goeree2016} for further discussion). As shown by \cite{luce1959individual}, logit is characterized by Independence of Irrelevant Alternatives (IIA), which states that the relative choice frequencies of two alternatives are independent of other alternatives. In other words, after taking account of the utilities of alternatives, all alternatives are treated equally. There are several well-documented violations of IIA, and violations of IIA are also documented in games. IIA is violated when some alternatives are salient or focal because salience or focality of alternatives usually cannot be fully explained by differences in utilities.   

We will now define the logistic version of the Focal QRE. The Focal Logit Equilibrium is a generalization of the Logit QRE that allows for behavioral forces, such as focality or salience, that are ruled out by IIA while still retaining attractive properties of logit.

\begin{defn}[Focal Logit Equilibrium] Let $\Gamma=(N, S, u)$ be a game in normal form. For any $\lambda\in \mathds{R}_{+}$, a \textbf{Focal Logit Equilibrium} for a given $(F_i, \delta_i)_{i\in N}$ is any $\pi\in\Delta$ such that for all $i\in N$ and $j\in J_i$,
\[\pi_{ij}=\frac{\exp\Big(\lambda\big(u_{i}(s_{ij}, \pi_{-i})+\delta_i(\Gamma)\mathds{1}\{s_{ij}\in F_i(\Gamma)\}\big)\Big)}{\sum_{k\in J_i}\exp\Big(\lambda\big(u_{i}(s_{ik}, \pi_{-i})+\delta_i(\Gamma)\mathds{1}\{s_{ik}\in F_i(\Gamma)\}\big)\Big)}.\] 
\end{defn}

Indeed, when $\delta_i(\Gamma)=0$ or $F_i(\Gamma)=S_i$, the Focal Logit Equilibrium reduces to the logistic QRE. In the following two sections, we apply the Focal Logit Equilibrium to some well-known games.

\subsection{Identifying Focal Strategies}\label{identify}

Before we proceed to applications of the Focal QRE, we will briefly discuss how to identify focal strategies from observed choice frequencies. Our result shows that, under focal logit equilibrium, we may identify focal strategies from equilibrium choice frequencies. Roughly speaking, if $s_i$ is chosen more frequently than $s'_i$ after taking account of their utilities, then $s_i$ is focal and $s'_i$ is nonfocal. 

\begin{prps}\label{Flogit} Suppose $\sigma^*$ is a Focal Logit Equilibrium of $\Gamma$. Consider players $i, j\in N$ and strategies $s_i, s'_i\in S_i$ and $s_j, s'_j\in S_j$.

\begin{itemize}
\item[i)] If $\sigma^*_i(s_i)>\sigma^*_i(s'_i)$ and $u_i(s_i, \sigma^*_{-i})\le u_i(s'_i, \sigma^*_{-i})$, then $s_i\in F_i(\Gamma)$ and $s'_i\not\in F_i(\Gamma)$.\medskip
\item[ii)] Suppose $u_i(s_i, \sigma^*_{-i})>u_i(s'_i, \sigma^*_{-i})$ and $u_j(s_j, \sigma^*_{-j})>u_j(s'_j, \sigma^*_{-j})$. Then 
\[\!\!\!\!\!\!\!\!\!\!\!\!\!\!\!\frac{\log(\frac{\sigma^*_i(s_i)}{\sigma^*_i(s'_i)})}{\log(\frac{\sigma^*_j(s_j)}{\sigma^*_j(s'_j)})}\!>\!\frac{u_i(s_i, \sigma^*_{-i})\!-\!u_i(s'_i, \sigma^*_{-i})}{u_j(s_j, \sigma^*_{-j})\!-\!u_j(s'_j, \sigma^*_{-j})}\!\Rightarrow s_i\in F_i(\Gamma), s'_i\not\in F_i(\Gamma)\text{ or }s_j\not\in F_j(\Gamma), s'_j\in F_j(\Gamma).\]
\end{itemize}
\end{prps}

For example, consider $\Gamma_5$ of \cite{holt2024bilateral} in \autoref{ADfreq}. By applying Proposition \autoref{Flogit}, we can conclude that $U$ is focal and $D$ is non-focal, which is consistent with our assumption in \autoref{hetero}. Similarly, consider the Game M1 of  \cite{schotter1994laboratory} in \autoref{fig:M1}. By applying Proposition \autoref{Flogit}, we can conclude that $U$ is focal and $D$ is non-focal, which is consistent with our assumption in \autoref{app:regret}.

\smallskip
\noindent\textbf{Comment on Focal Strategies.} The idea of focal strategies and focal points in games goes back to at least \cite{schelling1960}. He, along with the subsequent literature, point out that even a simple relabeling of strategies could drastically change the outcome of games. For instance, imagine a coordination game where players have to choose a meeting point in New York. Intuitively, ``Grand Central Station'' may be considerably more focal than many other locations. Although there are scenarios where they may coincide, focal strategies in our paper do not necessarily coincide with ones in the sense of \cite{schelling1960}.

In our framework, focality captures strategies that stand out for reasons that are not related to strategic concerns or utilities of choices. We described earlier that focality can arise from regret, saliency, and limited consideration/attention, and so our notion of focality is primarily intended to capture such forms of bounded rationality.  Still, our framework does not rule out a more nuanced, context-dependent focality, as discussed in \cite{schelling1960}.  Indeed, by introducing a richer description of the game (e.g., since the strategy labels have meaning), the notion of focality in our framework is consistent with the idea of Schelling. 

However, the literature on focal points typically tries to predict the most frequently played strategy profile, for example, from the payoff structures of games and \emph{the labeling of strategies}. Similarly, the idea of focal points is often used to make an equilibrium selection (see \cite{mehta1994} for discussion and an experimental study of \cite{schelling1960}). However, in our framework, there can be multiple focal strategies for a single player, and some of them might be chosen infrequently (for us, roughly speaking, a strategy is focal if it is chosen relatively more frequently than its expected utility suggests). Moreover, some strategies that are not part of any equilibrium profile can be focal. In that sense, focal strategies in our framework may not coincide with ones previously considered and studied in the literature.

One advantage of our framework is that we can sometimes identify focal strategies from observed frequencies without relying on any theory of focality.

\section{Across Game Heterogeneity}\label{hetero}

\subsection{Predictive Power of Focal QRE}

The ability to make predictions about one economic environment from similar economic environments is essential for economics as a science. Accordingly, it is often assumed that economic primitives, including preferences of economic agents (e.g., risk attitudes, discount rates), are stable across different environments. Indeed, the standard microeconomics course begins by introducing an economic agent with a stable preference making choices from different choice sets. Without stability of the economic primitives, the standard economic model loses a significant portion of its predictive power unless strong parametric assumptions are made. 
Many of the results in choice theory and revealed preference theory rely on stability of preferences (\cite{arrow1959rational}, \cite{afriat1967construction}, \cite{varian1982}). 

Utility maximization loses significant appeal without assuming stability of preferences or allowing for choice observations from different choice sets. The same can be said for the random utility model. Indeed, stability of the distribution of preferences (i.e., error term) is almost always assumed in choice theory and the revealed preference literature (\cite{falmagne1978representation}, \cite{mcfadden1990stochastic}, \cite{gul2006}, \cite{kitamura2018}). Hence, it is natural to assume stability of the distribution of preferences in the QRE. 

Assuming stability of the distribution of preferences is particularly reasonable in experimental settings, since observed choice frequencies are usually calculated from choices from the same set of subjects. For this reason, we follow this tradition and assume stability of the distribution of preferences, which is equivalent to stability of the error terms and risk preferences.     

When error terms are stable across similar games, the QRE has strong predictive power and it can even be rejected non-parametrically (\cite{Melo2019}). Indeed, QRE can be rejected non-parametrically by using observations from two 2$\times$2 games. As our applications show below, the logistic QRE can be rejected even from the choice frequencies of a single 2$\times $2 game.

Similarly, under a fixed error distribution, the Focal QRE can be rejected non-parametrically without any assumptions on error terms, utility function, and risk preferences. For example, it can be rejected by using observations from four 2$\times$ 2 games (See Appendix A.4). The Focal Logit Equilibrium makes even sharper predictions; it can be rejected with choice frequencies from two 2$\times$2 games or a single 2$\times$3 game (see applications below). In the next subsection, we show that the Focal Logit Equilibrium with a stable distribution of preferences (i.e., a stable error distribution) can explain across game heterogeneity observed in experiments.

In addition to the axiomatic characterization of the Focal Luce Model (i.e., the logit model with focality) appearing in \cite{kovach2017imbalanced}, we also formally show that because of the logit formulation and the fact that focal bias symmetrically affects all alternatives, the magnitude of focality and the focal set can be considered at most one free parameter. Therefore, for $n$-player normal form games, the Focal Logit Equilibrium has at most $n+1$ free parameters (ignoring risk preferences).  This is why the model can be rejected by choice frequencies from a single $2\times 3$ game.

The remainder of this paper focuses on the Focal Logit Equilibrium.

\subsection{Explaining Play Heterogeneity Across Games with Focal Logit Equilibrium}\label{het:FLE}

Since we have just argued that it is natural to impose a stable distribution of errors across games in the QRE and the Focal QRE, it is natural to wonder how well the QRE can explain play across similar games.  Somewhat problematically, the experimental game theory literature finds that observed play is close to Nash equilibrium in some games, but is far away from Nash equilibrium in others.  This difference persists even after adjusting for differences in payoffs (e.g., see \cite{goeree2001ten} and \cite{camerer2004cognitive}). 

For instance, if we fix the error distribution to be type I extreme value, then we have the logistic QRE and bounded rationality is captured by a single parameter ($\lambda$). Ideally, we would like the estimates of the QRE from observed choice frequencies to be consistent across games.  Unfortunately, estimates of $\lambda$ for the logistic QRE may differ by an order of magnitude across games. In the rest of this section we demonstrate that the Focal QRE can explain this heterogeneous play across games (via changes in the salient/focal strategies) while maintaining consistent estimates of $\lambda$, thereby explaining changes in play across games without assuming a change in the underlying distribution of preferences.

The failure of the (logistic) QRE to fully account for play across games should not be surprising. Indeed, as discussed in the introduction, there is overwhelming evidence from the literature on individual decision making that choice behavior cannot always be explained by (random) utility maximization (e.g., choice overload and the attraction effect).  Additionally, people even fail to perform ``soft" utility maximization because of (for example) salience, limited consideration, and reference dependence. Focal choice can systematically capture these different ideas of bounded rationality without resorting to arbitrary changes in the distribution of preferences, thereby retaining testable implications across games.

\subsubsection{(Asymmetric) Matching Pennies Games}

\begin{table}
\centering

\begin{minipage}{0.32\textwidth}
	\begin{tabular}{|c|c|c|c|} \hline
		$\Gamma_1$	& \textbf{\color{blue}L} & \textbf{\color{blue}R}  & \\ \hline
    	 \textbf{\color{blue}U} & 8, 4 & 4, 8 & 0.52\\ \hline
    	 \textbf{\color{blue}D} & 4, 8 & 8, 4 & 0.48\\ \hline
	 \hline
    			& 0.52 & 0.48 & Freq \\ \hline
 			& 0.5 & 0.5 & Focal \\ \hline
	 \end{tabular}
\end{minipage}
\begin{minipage}{0.32\textwidth}
	 \begin{tabular}{|c|c|c|c|} \hline
        		$\Gamma_2$	& \textbf{\color{blue}L} & \textbf{\color{blue}R}  & \\ \hline
    	 \textbf{\color{blue}U} & 32, 4 & 4, 8 & 0.96\\ \hline
    		    D & 4, 8 & 8, 4 & 0.04\\ \hline
		    \hline
     & 0.16 & 0.84 &  Freq \\ \hline
       & 0.17 & 0.83 &  Focal \\ \hline
 	 \end{tabular}
\end{minipage}
\begin{minipage}{0.32\textwidth}
	\begin{tabular}{|c|c|c|c|} \hline
        		   $\Gamma_3$	& \textbf{\color{blue}L} & \textbf{\color{blue}R}  & \\ \hline
		     U & 4.4, 4 & 4, 8 & 0.08\\ \hline
    	 \textbf{\color{blue}D} & 4, 8 & 8, 4 & 0.92\\ \hline
	 \hline
      			& 0.8 & 0.2 & Freq \\ \hline
			& 0.81 & 0.19 & Focal \\ \hline
 	\end{tabular}
\end{minipage}
\caption{Symmetric and Asymmetric Matching Pennies Games in Goeree and Holt (2001)}\label{mpfreq}
\end{table}

We illustrate how the Focal QRE can explain heterogeneity across games via three versions of matching pennies games from \cite{goeree2001ten}. The first game in \autoref{mpfreq} is a standard symmetric matching pennies game, while the second and third games in \autoref{mpfreq} are derived from the first one by asymmetrically changing Player 1's payoffs at $(U, L)$.\footnote{Each game that is presented as a payoff table is labeled on the left top corner of the matrix. The focal strategies are highlighted in blue.} With these three versions of the matching pennies game (along with other well-known games), \cite{goeree2001ten} demonstrate that observed behavior is close to Nash equilibrium in some games, but can be far away from Nash equilibrium in other, strategically-similar games. Notice that the observed behavior in the symmetric matching pennies game is very close to the choice frequencies predicted by Nash equilibrium. However, the observed behavior in the asymmetric games are highly inconsistent with Nash equilibrium. In fact, it is difficult to explain the two asymmetric matching pennies games simultaneously by the QRE with the same parameters.

The Focal Logit Equilibrium can explain this data. Suppose that $F_1(\Gamma_2)=\{U\}$, since $(U, R)$ returns the highest possible payoff $\$ 3.20$, and that $F_2(\Gamma_2)=\{L, R\}$, since $L$ and $R$ are relatively symmetric. However, in $\Gamma_3$, suppose that $F_1(\Gamma_3)=\{D\}$ since $D$ almost dominates $U$ and $F_2(\Gamma_3)=\{L, R\}$ since $L$ and $R$ are relatively symmetric. The proposed focal sets are consistent with the idea of salience as well as regret aversion (See Observation \ref{obs4} and section \ref{mpg}). Then a strategy profile $((p_1, 1-p_1), (q_1, 1-q_1))$ in $\Gamma_2$ is a Focal Logit Equilibrium if
\begin{equation}
\frac{p_1}{1-p_1}=\frac{\exp\!\Big(\lambda\,\big(32\,q_1+4\,(1-q_1)+\delta(\Gamma_2)\big)\Big)}{\exp\!\Big(\lambda\, \big(4\,q_1+8\,(1-q_1)\big)\Big)}\end{equation}
and
\begin{equation}
\frac{q_1}{1-q_1}=\frac{\exp\!\Big(\lambda\,\big(4\,p_1+8\,(1-p_1)\big)\Big)}{\exp\!\Big(\lambda\,\big(8\,p_1+4\,(1-p_1)\big)\Big)}.\end{equation}
By direct calculations, the observed frequencies $((0.96, 0.04), (0.16, 0.84))$ are consistent with the Focal Logit equilibrium when $\lambda=0.45$ and $\delta(\Gamma_2)=5.93$. 

Similarly, a strategy profile $((p_2, 1-p_2), (q_2, 1-q_2))$ in $\Gamma_3$ is a Focal Logit Equilibrium if
\begin{equation}
\frac{p_2}{1-p_2}=\frac{\exp\!\Big(\lambda\,\big(4.4\,q_2+4\,(1-q_2)\big)\Big)}{\exp\!\Big(\lambda\,\big(4\,q_2+8\,(1-q_2)+\delta(\Gamma_3)\big)\Big)}\end{equation}
and
\begin{equation}
\frac{q_2}{1-q_2}=\frac{\exp\!\Big(\lambda\,\big(4\,p_2+8\,(1-p_2)\big)\Big)}{\exp\!\Big(\lambda\,\big(8\,p_2+4\,(1-p_2)\big)\Big)}.\end{equation}

By direct calculations, the observed frequencies $((0.08, 0.92), (0.8, 0.2))$ are consistent with the Focal Logit equilibrium when $\lambda=0.41$ and $\delta(\Gamma_3)=5.44$.\footnote{Interestingly, the column players are very consistent with the QRE. In fact, $\lambda$ calculated from $(8)$ and $(10)$ are very similar, $0.41$ and $0.45$. Therefore, once the row players' behavior and focal strategies ($U$ in $\Gamma_2$ and $D$ in $\Gamma_3$) are accounted for, the observed frequencies are highly consistent with the QRE.} Since the calculated parameters are different ($\lambda=0.45$ and $\lambda=0.41$), we cannot perfectly replicate the data with the Focal Logit equilibrium. However,  $0.45$ and $0.41$ are very similar, and thus we can explain most of the heterogeneity across the games. Interestingly, the focality bias terms are also quite similar across games, $5.93$ and $5.44$.  
 
However, in the logistic QRE (i.e., $\delta(\Gamma_2)=\delta(\Gamma_3)=0$), $(7)$ and $(8)$ imply that $\lambda=2.84$ and $\lambda=0.45$, respectively, while $(9)$ and $(10)$ imply that $\lambda=5.09$ and $\lambda=0.41$, respectively.\footnote{We wish to note that the difference between $\lambda=2.84$ and $\lambda=0.45$ can be explained by introducing risk-loving preferences over payoffs (CRRA parameter is $\gamma=1.7$ where $u(x)=x^\gamma$).} However, it is difficult to explain the difference between $\lambda=5.09$ and $\lambda=0.41$ with standard arguments. In fact, $\Gamma_3$ cannot be explained by any logistic QRE. Indeed, this precisely illustrates the heterogeneity of rationality across games (even in strategically-similar games). The Focal QRE can account for a large portion of observed heterogeneity by incorporating salience and regret aversion.

The calculations above solve for game specific point-values of $\lambda$ to illustrate changes in play across games. We now complement those calculations with a maximum-likelihood exercise that estimates a common parameter across
games. This exercise is useful for two reasons. First, it is close to the standard empirical implementation of Logit QRE. Second, it clarifies the extent to which Focal Logit improves fit after accounting for the number of additional parameters. 

The results, shown in \autoref{tab:mle-matching-pennies}, compares three specifications: standard Logit QRE with a common precision
parameter \(\lambda\), Focal Logit with common \((\lambda,\delta)\), and Focal
Logit with common \(\lambda\) but game-specific focality biases
\((\delta_2,\delta_3)\). For the focal specifications, we use the focal sets derived previously: $F^R_{\Gamma_2}=\{U\}$ and $F^R_{\Gamma_3}=\{D\}$.

\begin{table}[ht]
\centering
\footnotesize
\setlength{\tabcolsep}{3.5pt}
\begin{tabular}{llccc}
\toprule
Model & Estimates & Game & Obs./Pred. \((p,q)\) & AIC/BIC \\
\midrule
Logit QRE
&
\(\hat\lambda=.629\)
&
\(\Gamma_2\)
&
\((.96,.16)/(.788,.190)\)
&
89.34/91.25
\\
&
&
\(\Gamma_3\)
&
\((.08,.80)/(.348,.682)\)
&
\\
\midrule
Focal Logit,
common \(\delta\)
&
\(\hat\lambda=.423,\ \hat\delta=5.496\)
&
\(\Gamma_2\)
&
\((.96,.16)/(.954,.177)\)
&
73.43/77.25
\\
&
&
\(\Gamma_3\)
&
\((.08,.80)/(.075,.808)\)
&
\\
\midrule
Focal Logit,
game-specific \(\delta_g\)
&
\(\hat\lambda=.433\)
&
\(\Gamma_2\)
&
\((.96,.16)/(.961,.168)\)
&
75.37/81.11
\\
&
\(\hat\delta_2=6.019,\ \hat\delta_3=5.128\)
&
\(\Gamma_3\)
&
\((.08,.80)/(.082,.809)\)
&
\\
\bottomrule
\end{tabular}

\vspace{0.5em}
\caption{MLE estimates for asymmetric matching-pennies games. In the table, \(p\) is the row player's probability of choosing \(U\), and \(q\) is the column player's probability of choosing \(L\). In the Obs./Pred. column, the first pair is observed and the second is predicted.}
\label{tab:mle-matching-pennies}

\end{table}

The common-\(\lambda\) Logit QRE captures the column frequencies reasonably well,
but it misses the large row-player shift across the two asymmetric games. Adding a
single common focality-bias parameter substantially improves fit: the common-\(\delta\)
Focal Logit model has AIC $73.43$ and BIC $77.25$, compared with $89.34$ and $91.25$ for standard Logit
QRE.

\subsubsection{Attacker-Defender Games}

As a second illustration, we now consider the attacker-defender games studied experimentally by \cite{holt2024bilateral}. The attacker-defender games, shown in \autoref{ADfreq}, are 2$\times$2, zero-sum games which are strategically similar to the asymmetric matching pennies games. \cite{holt2024bilateral} found that when payoffs from strategies $U$ and $D$ are relatively close (i.e., in the left-panel version -- $\Gamma_4$), the observed frequencies are very close to Nash equilibrium predictions. However, when payoffs from strategies $U$ and $D$ are relatively distinct (i.e., in the middle-panel and right-panel versions -- $\Gamma_5$ and $\Gamma_6$), observed frequencies depart significantly from NE frequencies. 

\begin{table}[h]
\begin{minipage}{0.31\textwidth}
	\begin{tabular}{|c|c|c|c|} \hline
		$\Gamma_4$	& \textbf{\color{blue}L} & \textbf{\color{blue}R}  & \\ \hline
    	 \textbf{\color{blue}U} & -3, 3 & -7, 7 & 0.34 \\ \hline
    	 \textbf{\color{blue}D} & -6, 6 & -4, 4 & 0.66 \\ \hline
	 \hline
    			& 0.49 & 0.51 & Freq \\ \hline
 	 \end{tabular}
\end{minipage}
\begin{minipage}{0.31\textwidth}
	 \begin{tabular}{|c|c|c|c|} \hline
        		$\Gamma_5$	& L & \textbf{\color{blue}R}  & \\ \hline
    	 \textbf{\color{blue}U} & -2, 2 & -7, 7 & 0.59  \\ \hline
    		  D & -8, 8 & -4, 4 & 0.41  \\ \hline
		    \hline
      & 0.33 & 0.67 &   Freq \\ \hline
 	 \end{tabular}
\end{minipage}
\begin{minipage}{0.31\textwidth}
	\begin{tabular}{|c|c|c|c|c|} \hline
        		   $\Gamma_6$	& \textbf{\color{blue}L} & R  & \\ \hline
		     \textbf{\color{blue}U} & -6, 6 & -7, 7 & 0.82  \\ \hline
    	 D & -12, 12 & -4, 4 & 0.18 \\ \hline
	 \hline
      		& 0.61 & 0.39 & Freq \\ \hline
 	\end{tabular}
\end{minipage}
\caption{Attacker-defender games in \cite{holt2024bilateral}}\label{ADfreq}
\end{table}

Similar to the previous exercise, let us first derive $\lambda$, assuming the Logistic QRE. The choice frequencies of the row player in $\Gamma_5$ imply that $\lambda=0.54$ for the column player, but in $\Gamma_6$ the choice frequencies of the row and column players imply $\lambda=0.7$ and $\lambda=0.54$, respectively. However, the choice frequencies of the column player in $\Gamma_5$ cannot be explained by any Logistic QRE (not even any Symmetric QRE) under risk neutrality. That is because symmetric QRE implies that $D$ must be chosen more frequently than $U$, since $0.33\times 2+0.67\times 7>0.33\times 8+0.67\times 4$. When risk-aversion is introduced, assuming CRRA preferences, $\lambda=18$ under extreme risk aversion. All choice frequencies can be matched very well with the Focal Logit Equilibrium with the same $\lambda=0.41$.

\subsection{Risk Attitudes}

In experiments, payoffs are written in terms of monetary rewards, which may not correspond exactly to the players' utilities. Because of this, it makes sense to consider players' risk attitudes. While introducing risk attitudes can improve fit, this does not typically solve the problem of across game heterogeneity. In fact,  accommodating the observed changes in play across strategically similar games requires dramatic changes in risk attitudes, both quantitatively and qualitatively. For example, in the matching pennies games above, assuming the logistic QRE, we find that $\Gamma_2$ implies moderate risk-seeking behavior, but $\Gamma_3$ is not consistent with any logistic QRE under expected utility theory. Similarly, in the attacker-defender games above, $\Gamma_5$ implies extreme risk-aversion, but $\Gamma_6$ would imply either extreme risk-seeking or risk-averse behavior. In the coordination games with a secure outside option studied in \autoref{sec:regret}, we find that $\Gamma_1$ implies moderate risk-aversion, while $\Gamma_2$ implies risk neutrality.  For strategically similar games, it is implausible that risk attitudes of the population change this sharply. 

The advantage of the Focal Logit QRE is that it can explain play in the aforementioned games while maintaining stability of the underlying preferences.  Following the preference heterogeneity interpretation of RUM, the measured risk attitude should approximate the population average, and so ``risk neutrality'' under the QRE means that the subjects are on average close to risk neutral. Since the Focal QRE can fit the data without requiring changes in risk attitudes, it is consistent with a stable underlying distribution of preferences.

\section{Modeling Focality}\label{sec:regret}

The Focal QRE is quite useful when explaining behavior that is inconsistent with the standard QRE, especially when reconciling play across strategically similar games. These findings do not rely on assumptions about how focality was determined. Similarly, our results on falsifiability of the Focal QRE do not rely on assumptions about focality.  This generality allows the Focal QRE to serve as a structural framework to test theories of bounded rationality in games. We illustrate this by introducing an explicit theory of focality based on regret aversion. This allows focal sets to be constructed directly from the payoff structure before observing choice frequencies and enables the Focal QRE to make considerably sharper predictions,

\subsection{Regret-Averse Focal Sets}

In this section, we consider focal sets that are composed of strategies with low ex-post regret. This specification is behaviorally meaningful and, since focal sets are uniquely determined from the game's payoffs, it imposes strong restrictions on the data. Accordingly, only the bias term remains as a free parameter. Thus the Focal QRE with regret-aversion is a one-parameter extension of the QRE.

 We then apply this specification to the five normal-form games from \cite{goeree2001ten}.  Each of the five games are of special interest to economists because subjects' play in experiments is typically far from equilibrium predictions. More importantly, the deviations from equilibrium  in each game are typically explained with a different auxiliary theory. We show that play in some games is quite consistent with regret-averse focal sets.

In order to specify regret-averse focal sets, we need to establish a bit of notation. For a given $s\in S$, let $\max_{s'_i\in S_i\setminus s_i} u_i(s'_i, \textbf{s}_{-i})-u_i(s_i, \textbf{s}_{-i})$ be the amount of (ex-post) regret that player $i$ could experience playing $s_i$. Then for each $s_i$, player $i$ may anticipate the maximum regret level $R_i(s_i)\equiv \max_{\textbf{s}_{-i}\in S_{-i}}\big(\max_{s'_i\in S_i\setminus s_i} u_i(s'_i, \textbf{s}_{-i})-u_i(s_i, \textbf{s}_{-i})\big)$.

\begin{defn}[Regret-Averse Focal Set] For a given normal-form game $\Gamma$, let $\bar{R}_i(\Gamma)\equiv \frac{\sum_{s_i\in S_i}R_i(s_i)}{|S_i|}$ be the average maximum regret level for player $i$. Then the \emph{regret-averse focal set} for player $i$ in $\Gamma$ is defined by
\begin{equation}\label{eq:regret}F_i(\Gamma)=\{s_i\in S_i: R_i(s_i)\le \bar{R}_i(\Gamma)\}.\end{equation}\end{defn}

Regret-averse focal sets have several desirable properties, which we illustrate in the following three observations. First, dominant strategies are always focal.

\begin{obs}\label{obs1} $s_i\in F_i(\Gamma)$ if at least one of the following two conditions is satisfied.

\begin{enumerate} 
\item[i)] $s_i\in S_i$ is a weakly dominant strategy for player $i$;
\item[ii)] $s_i$ weakly dominates $s'_i$ and $s'_i\in F_i(\Gamma)$.
\end{enumerate}
\end{obs}

Strategies that return the highest payoff are not necessarily in focal sets, as they might return very low payoffs in other instances.  The following observation shows that if $s_i$ returns the highest payoff, and the highest payoff is significantly larger than other payoffs, then $s_i$ is focal. Similarly, if the lowest payoff is significantly lower than other payoffs, then the strategy returning the lowest payoff is non-focal.

\begin{obs}\label{obs2}  Let $u_i(\bar{\textbf{s}})=\max_{s\in S} u_i(s)$ and $u_i(\underline{\textbf{s}})=\min_{s\in S} u_i(s)$. 
\begin{enumerate}
\item [i)] If $u_i(\bar{\textbf{s}})\ge 2\max_{s\in S\setminus \bar{\textbf{s}}} u_i(s)-u_i(\underline{\textbf{s}})$, then $\bar{s}_i\in F_i(\Gamma)$. 
\item [ii)] If $u_i(\underline{\textbf{s}})<2\min_{s\in S\setminus \bar{\textbf{s}}} u_i(s)-u_i(\bar{\textbf{s}})$, then $\underline{s}_i\not\in F_i(\Gamma)$.
\end{enumerate}
\end{obs}

Observation \autoref{obs2} is useful when checking whether a strategy is focal. For example, in the first asymmetric matching pennies game (see \autoref{mpfreq}), $U$ returns the highest payoff $32$, which is greater than $12=2\cdot 8-4$. Therefore, $U$ is a focal strategy under regret-aversion.

Intuitively, a strategy $s_i$ leads to regret when an agent ends up with a much lower payoff than she could have. If $s_i$ features relatively little payoff variation and a high average payoff, we would therefore expect the level of regret associated with $s_i$ to be low. In other words, if $s_i$ returns high payoffs regardless of other players' strategies, then $s_i$ is focal. Observation  \autoref{obs3} formalizes this.
 
\begin{obs}\label{obs3} Let $u_i(\bar{\textbf{s}})=\max_{s\in S} u_i(s)$. Then $s_i\in F_i(\Gamma)$ if at least one of the following two conditions is satisfied.
\begin{enumerate} 
\item[i)] $\min_{s_{-i}\in S_{-i}}u_i(s_i, s_{-i})\ge \frac{1}{2}\,u_i(\bar{s})+\frac{1}{2\,(|S_i|-1)}\big(\sum_{s'_i\neq s_i }\min_{s_{-i}\in S_{-i}}u_i(s'_i, s_{-i})\big)$;
\item[ii)] $u_i(s_i, s_{-i})\ge\frac{u_i(\bar{s})+\sum_{s'_i\neq s_i }u_i(s'_i, s_{-i})}{|S_i|}$ for each $s_{-i}\in S_{-i}$.
\end{enumerate}
\end{obs}

Finally, we show that for $2\!\times\!2$ games it is sufficient to calculate average payoffs to identify the focal strategies. 

\begin{obs}\label{obs4} Suppose $n=2$ and $S_1=\{s^1_1, s^2_1\}$ and $S_2=\{s^1_2, s^2_2\}$. Then $s_i\in F_i(\Gamma)$ if and only if
\[u_i(s_i, s^1_j)+u_i(s_i, s^2_j)\ge u_i(s'_i, s^1_j)+u_i(s'_i, s^2_j).\]
\end{obs}

\subsection{Applying Regret-Averse Focal Sets}\label{app:regret}

We now apply the focal logit equilibrium with regret-averse focal sets. Consider the following two-player, normal-form game studied in \cite{schotter1994laboratory}. Strategies and payoffs as well as observed frequencies of strategies are depicted in \autoref{fig:M1}. Notice that this game has two Nash equilibria, $(U, L)$ and $(D, R)$. However, in the experimental study of \cite{schotter1994laboratory}, the observed frequency of $(U, R)$ is the highest (with probability $0.456=0.8\times 0.57$) among all four pure strategy profiles.

\begin{table}
\centering

\begin{tabular}{|c|c|c|c|} \hline
	$\Gamma$		& L & \textbf{\color{blue}R}  & \\ \hline
    	 \textbf{\color{blue}U} & 4, 4 & 4, 4 & 0.57 \\ \hline
    	 D & 0, 1 & 6, 3 & 0.43\\ \hline
	 \hline
    			& 0.2 & 0.8 & Freq \\ \hline
 	 \end{tabular}

\caption{Game M1 of \cite{schotter1994laboratory}}\label{fig:M1}
\end{table}

By Observation \autoref{obs4}, $F_1(\Gamma)=\{U\}$ since $4+4>0+6$ and $F_2(\Gamma)=\{R\}$ since $4+3>4+1$. Intuitively, $U$ is focal since $U$ returns $\$ 4$ for sure and $R$ is focal since $R$ weakly dominates $L$ for Player 2.

Therefore, a strategy profile $((p, 1-p), (q, 1-q))$ in $\Gamma$ is a \textbf{Focal Logit Equilibrium} if
\[\frac{p}{1-p}=\frac{\exp\!\Big(\lambda\,\big(4+\delta(\Gamma)\big)\Big)}{\exp\!\Big(\lambda\,\big(6\,(1-q)\big)\Big)}\text{ and }\frac{q}{1-q}=\frac{\exp\!\Big(\lambda\,\big(4\,p+(1-p)\big)\Big)}{\exp\!\Big(\lambda\,\big(4\,p+3\,(1-p)+\delta(\Gamma)\big)\Big)}.\]
By direct calculations, the observed frequencies $((0.57, 0.43), (0.2, 0.8))$ are consistent with the Focal Logit Equilibrium when $\lambda=0.66$ and $\delta(\Gamma)=1.22$. However, when $\delta(\Gamma)=0$ (i.e., the logistic QRE), there is no $\lambda$ that can generate these frequencies.\footnote{In fact, these choice frequencies cannot be explained by the QRE without introducing risk aversion. The reason is as follows. Given that $R$ is chosen with probability $0.8$, $D$'s expected payoff of $4.8$ is higher than $U$'s expected payoff of $4$. Therefore, by Property 4 of the quantal response function, $\sigma$, in any QRE $D$ must be chosen with probability at least $0.5$.   The QRE with moderate risk aversion can explain this data, however.}

\subsection{Five Treasures of Game Theory}
 
In this subsection, we apply the theory of regret-averse focal sets to the five normal-form games in \cite{goeree2001ten}.
 
\subsubsection{Matching Pennies Games}\label{mpg}

The first normal-form games of \cite{goeree2001ten} are the matching pennies games we described in \autoref{het:FLE}. Let us derive regret-averse focal sets for the three matching pennies games in \autoref{mpfreq}. Since all three are $2\times 2$ games, we can apply Observation \autoref{obs4}.

In the symmetric matching pennies game ($\Gamma_1$ of \autoref{mpfreq}), both strategies are included in the regret-averse focal set for each player. More precisely, by Observation \autoref{obs4} we have $F_1(S_1)=\{U, D\}$ and $F_2(S_2)=\{L, R\}$ since $8+4=4+8$. Therefore, the Focal QRE and the QRE coincide. Moreover, in both asymmetric matching pennies games, $\Gamma_2$ and $\Gamma_3$, $F_2(S_2)=\{L, R\}$ since the column players' strategies are payoff symmetric (i.e., $8+4=4+8$). 

In the first asymmetric matching pennies game, $\Gamma_2$, player one's payoff in $(U, L)$ is significantly increased from $8$ to $32$, and therefore $F_1(S_1)=\{U\}$ since $32+4>4+8$ (see also Observation \autoref{obs2}). On the other hand, in the second asymmetric matching pennies game, $\Gamma_3$, player one's payoff in $(U, L)$ is only $4.4$, and therefore $F_1(S_1)=\{D\}$ since $4.4+4<4+8$. Intuitively, $U$ is focal in $\Gamma_2$ since $(U, L)$ returns the highest possible payoff $\$ 3.20$, while $D$ is focal in $\Gamma_3$ since $D$ almost dominates $U$.

\begin{table}
\centering
     \begin{subtable}[b]{0.48\textwidth}
\centering
    \resizebox{\columnwidth}{!}{
\begin{tabular}{ |c | c | c | c | c | c|}
\hline
$\Gamma_1$ &  Left  & {\color{blue}{\textbf{Right}}}  & Safe &  \\ \hline
  U & 9, 9 & 0, 0 & 0, 4 &  0.04\\ \hline
    {\color{blue}{\textbf{D}}} & 0, 0 & 18, 18& 0, 4 &  0.96 \\ \hline
      &  ? &  0.84 & ? & Freq \\ \hline
  \end{tabular}}
     \end{subtable}
         \begin{subtable}[b]{0.48\textwidth}
\centering
    \resizebox{\columnwidth}{!}{
\begin{tabular}{ |c | c | c | c | c | c|}
\hline
$\Gamma_2$ &  Left  &  {\color{blue}{\textbf{Right}}}   & Safe &  \\ \hline
    {\color{blue}\textbf{U}} & 9, 9 & 0, 0 & 40, 4 &  0.36\\ \hline
    D & 0, 0 & 18, 18& 0, 4 &  0.64\\ \hline
     & ? &  0.76 & ?  & Freq\\ \hline
  \end{tabular}}
  \end{subtable}
\caption{A Coordination Game with a Secure Outside Option in \cite{goeree2001ten}}\label{fig:coord}
\end{table}

Therefore, the regret-based focal sets are consistent with observed choice frequencies in these matching pennies games. In fact, the observed choice frequencies can be matched relatively well with the Focal Logit Equilibrium with the same $\lambda=0.41$, as we demonstrated in Section 4.2.

\subsubsection{A Coordination Game with a Secure Outside Option}

The second normal-form game of \cite{goeree2001ten} is a coordination game with a secure outside option. In particular, \cite{goeree2001ten} consider two versions illustrated in \autoref{fig:coord}. In addition to standard coordination games, the column player can take a safe action, Safe, which returns 4 for sure. In the first version (i.e., $\Gamma_1$), Safe returns 0 for sure to the row player, while in the second version (i.e., $\Gamma_2$) Safe returns either 40 or 0 to the row player.\footnote{Choice frequencies of Left and Safe are not reported in \cite{goeree2001ten}.}

In both games, $R_2(\text{Right})=9$, $R_2(\text{Left})=18$, and $R_2(\text{Safe})=14$. Therefore, $F_2(\Gamma_1)=F_2(\Gamma_2)=\{\text{Right}\}$. Moreover, when $u_1(\text{U}, \text{Safe})=0$, since $R_1(\text{U})>R_1(\text{D})$, we have $F_1(\Gamma_1)=\{\text{D}\}$. However, when $u_1(\text{U}, \text{Safe})=40$, since $R_1(\text{U})=18<R_1(\text{D})=40$, we have $F_1(\Gamma_2)=\{\text{U}\}$. Therefore, regret-based focal sets can explain why row players choose U more frequently when $u_1(\text{U}, \text{Safe})=40$.

The logistic QRE implies four different possible parameter values: $\lambda=0.16$ and $\lambda\in (0.21, 0.23)$ from $\Gamma_1$ and $\lambda=0.23$ and $\lambda\in (0.05, 0.14)$ from $\Gamma_2$. However, the observed choice frequencies can be matched with the Focal Logit Equilibrium with the same $\lambda=0.15$.

\subsubsection{The Kreps Game}

The Kreps game is illustrated in \autoref{fig:kreps}. Of the column player's pure strategies, Non-Nash is the only one that is not part of any Nash equilibrium. However, \cite{goeree2001ten} found that around two-thirds of players select Non-Nash. To show that choosing Non-Nash is not a consequence of loss-aversion, they also consider a variation of the Kreps game where 30 (i.e., \$3) is added to all payoffs to avoid losses. The frequency with which players choose Non-Nash does not change significantly. 

\begin{table}
\centering
     \begin{subtable}[b]{0.50\textwidth}
\centering
    \resizebox{\columnwidth}{!}{
  \begin{tabular}{ |c | c | c | c | c | c | }
  \hline
    $\Gamma_1$  &  Left  & {\color{blue}{\textbf{Middle}}} & {\color{blue}{\textbf{Non-Nash}}} & Right & \\ \hline
    {\color{blue}{\textbf{U}}} & 20, 5 & 0, 4.5 & 1, 3 & 2, -25 & 0.68 \\  \hline
    D & 0, -25 & 1, -10& 3, 3 & 5, 4 & 0.32\\  \hline
 & 0.25 & 0.08 & 0.67 & 0 & Freq\\   \hline
  \end{tabular}}
  \end{subtable}
       \begin{subtable}[b]{0.48\textwidth}
    \resizebox{\columnwidth}{!}{
  \begin{tabular}{ |c | c | c | c | c | c | }
  \hline
    $\Gamma_2$  &  Left   & {\color{blue}{\textbf{Middle} }} & {\color{blue}{\textbf{Non-Nash}}} & Right & \\  \hline
    {\color{blue}{\textbf{U}}} & 50, 35 & 30, 34.5 & 31, 33 &32, 5 & 0.84\\ \hline
    D & 30, 5 & 31, 20& 33, 33 & 35, 34 & 0.16\\ \hline
      &  0.24 &  0.12 &  0.64 &  0 & Freq \\ \hline
  \end{tabular}}
  \end{subtable}
\caption{The Kreps Game in \cite{goeree2001ten}.}\label{fig:kreps}
\end{table}

In both versions, $F_1(\Gamma_1)=F_1(\Gamma_2)=\{U\}$ since $R_1(\text{U})=3<R_1(\text{D})=20$. Moreover, we can directly calculate that $R_2(\text{Right})=\text{30}>R_2(\text{Left})=\text{29}>\frac{30+29+14+2}{4}=18.75>R_2(\text{Middle})=14>R_2(\text{Non-Nash})=2$. Therefore, $F_2(\Gamma_1)=F_2(\Gamma_2)=\{\text{Middle}, \text{Non-Nash}\}$, which is consistent with the experimental finding that Non-Nash is chosen with a high frequency. However, regret-based focality is inconsistent with the fact that Left is chosen more frequently than Middle. 

According to this theory of regret-averse focal sets, the two versions of the Kreps game have the same focal sets. However, since the size of the bias term can depend on payoffs, in a Focal QRE the two versions can have slightly different choice frequencies.

\subsubsection{The Traveler's Dilemma Game}

In the traveler's dilemma game, both players pick natural numbers $n_1, n_2\in \{180, \ldots, 300\}$. Both players are paid the lower of the two numbers and, in addition, an amount $T$ is transferred from the player with the higher number to the player with the lower number. That is,
\[u_i(n_i, n_j)=\min\{n_i, n_j\}+T\,\text{sign}\{n_j-n_i\}.\]
\cite{goeree2001ten} consider two cases,  $T=180$ or $5$, and find that when $T=180$ most people choose around $180$.  In contrast, when $T=5$, most people choose around $300$. 

Although the Traveler's dilemma (and the following game) have a large number of strategies, we can still calculate regret-based focal sets. It turns out that when $T=180$, $R_i(180)=119$, $R_i(181)=180$, and $R_i(n_i)=359$ when $n_i>181$. In other words, when $T=180$, $F_i(\Gamma_1)=\{180, 181\}$.\footnote{$180$ becomes uniquely focal under regret minimization rather than below average regret focal sets.} However, when $T=5$, $R_i(n_i)=\max\{299-n_i, 9\}$ and $F_i(\Gamma_2)=\{240, 241, \ldots, 300\}$. Therefore, regret-based focal sets capture the switching behavior from $180$ to $300$ as $T$ decreases from $180$ to $5$.

\subsubsection{A Minimum-Effort Coordination Game}

The fifth and final normal-form game in \cite{goeree2001ten} is a minimum-effort coordination game. In this game, both players simultaneously choose integer effort levels $e_1, e_2\in [110, 170]$ and receive payoffs given by
\[u_i(e_i, e_j)=\min\{e_i, e_j\}-c\cdot e_i.\]
\cite{goeree2001ten} consider low and high cost treatments (i.e. $c=0.1$ or $0.9$) and find that players coordinate well in the low cost treatment (i.e., effort is $170$). In contrast, they fail to coordinate and choose $110$ about $50$ percent of the time in the high cost treatment.

It turns out that $R_i(e_i)=\max\{c\,(e_i-110), (1-c)(170-e_i)\}$. Therefore, when $c=0.9$, $F_i(\Gamma_1)=[110, 140]$ while when $c=0.1$, $F_i(\Gamma_2)=[140, 170]$. Therefore, regret-averse focal sets capture the switching behavior from $110$ to $170$ as cost changes from  $c=0.9$ to $c=0.1$.

\subsection{Discussion}

Regret-aversion is intuitive and has long been studied in the context of individual choice (\cite{loomes1982regret}). It is quite natural to think it plays a role in strategic settings as well, and the Focal QRE provides a tractable way to incorporate this behavior into games. The formula in \autoref{eq:regret}, while simple and often predictive, is not without its weaknesses. In the rest of this section we discuss some of the limitations of our specification and suggest some alternatives. 

First, since our formula depends on the average regret, it may be too permissive and yield implausibly large focal sets. For example, our notion of regret-aversion predicts that Middle is focal in the Kreps Game (\autoref{fig:kreps}). A one-parameter extension of \autoref{eq:regret}, so that $F_i(\Gamma)=\{s_i\in S_i: R_i(s_i)\le \beta\, \bar{R}_i(\Gamma)\}$ for $\beta \in (0,1]$, could partially address this weakness. If $\beta$ is not too large, then Middle is excluded. In this case, $\beta$ serves as a  ``permissiveness'' threshold, so that a strategy must have a substantially lower possible regret to be focal.

Second, regret-aversion only considers a player's own payoffs when determining focality. Since this is a strategic environment, simple introspection suggests that this restriction is likely to be violated. Additionally, the large literature on social preferences also suggests that this might be too restrictive. For instance, one might expect that strategies that yield Pareto efficient outcomes or outcomes that maximize the total payoff will be focal.\footnote{Fudenberg and Liang (2019) train algorithms to predict initial play in random games and find that being part of a Pareto Dominant NE is very predictive of play.} Such a specification would suggest that Left is focal in the Kreps Game (\autoref{fig:kreps}), which is in fact consistent with its high frequency of play. 

Third, there are various compelling alternatives to regret. For instance, the salience theory of \cite{bordalo2013salience} is one such example. Indeed, their salience theory can be considered a special case of general Regret Theory where the regret level for each action is calculated using a different formula from ours. 

Alternatively, focality or saliency of strategies can be determined using the Hurwicz criterion \citep{hurwicz1951}; i.e., one might presume that strategies with high Hurwicz $\alpha$-max-min values are focal: $F_i(\Gamma)= \{s_i \in S_i: H(s_i)\ge \bar{H}\}$ where $H(s_i)=\alpha \max_{\mathbf{s}_{-i}} u_i(s_i, \mathbf{s}_{-i})+(1-\alpha)\min_{\mathbf{s}_{-i}} u_i(s_i, \mathbf{s}_{-i})$ and $\alpha$ is the level of optimism and $\bar{H}=\sum_{s_i\in S_i} H(s_i)/J_i$.

\appendix
\section{Proofs}

\subsection{Proof of Proposition 1} (i). If $\sigma$ is a Focal Logit Equilibrium of $\Gamma$, then we have \[\log(\frac{\sigma_i(s_i)}{\sigma_i(s'_i)})=\lambda\big(u_i(s_i, \sigma_{-i})-u_i(s'_i, \sigma_{-i})+\delta_i(\Gamma)(\mathds{1}\{s_i\in F_i(\Gamma)\}-
\mathds{1}\{s'_i\in F_i(\Gamma)\})\big).\]
Hence, $\sigma(s_i)>\sigma(s'_i)$ and $u_i(s_i, \sigma_{-i})\le u_i(s'_i, \sigma_{-i})$ imply $\delta_i(\Gamma)(\mathds{1}\{s_i\in F_i(\Gamma)\}-
\mathds{1}\{s'_i\in F_i(\Gamma)\})>0$; equivalently, $s_i\in F_i(\Gamma)$ and $s'_i\not\in F_i(\Gamma)$.\smallskip

\noindent (ii). Since $\sigma^*$ is a Focal Logit Equilibrium of $\Gamma$, we have 
\[\log(\frac{\sigma_i(s_i)}{\sigma_i(s'_i)})=\lambda\big(u_i(s_i, \sigma_{-i})-u_i(s'_i, \sigma_{-i})+\delta_i(\Gamma)(\mathds{1}\{s_i\in F_i(\Gamma)\}-
\mathds{1}\{s'_i\in F_i(\Gamma)\})\big)\]
and 
\[\log(\frac{\sigma_j(s_j)}{\sigma_j(s'_j)})=\lambda\big(u_j(s_j, \sigma_{-j})-u_j(s'_j, \sigma_{-j})+\delta_j(\Gamma)(\mathds{1}\{s_j\in F_j(\Gamma)\}-
\mathds{1}\{s'_j\in F_j(\Gamma)\})\big).\]
Therefore, 
\[\frac{\log(\frac{\sigma_i(s_i)}{\sigma_i(s'_i)})}{\log(\frac{\sigma_j(s_j)}{\sigma_j(s'_j)})}=\frac{u_i(s_i, \sigma_{-i})-u_i(s'_i, \sigma_{-i})+\delta_i(\Gamma)(\mathds{1}\{s_i\in F_i(\Gamma)\}-\mathds{1}\{s'_i\in F_i(\Gamma)\})}{u_j(s_j, \sigma_{-j})-u_j(s'_j, \sigma_{-j})+\delta_j(\Gamma)(\mathds{1}\{s_j\in F_j(\Gamma)\}-\mathds{1}\{s'_j\in F_j(\Gamma)\})}.\]

Since $u_i(s_i, \sigma_{-i})>u_i(s'_i, \sigma_{-i})$ and $u_j(s_j, \sigma_{-j})>u_j(s'_j, \sigma_{-j})$ and $\frac{\log(\frac{\sigma_i(s_i)}{\sigma_i(s'_i)})}{\log(\frac{\sigma_j(s_j)}{\sigma_j(s'_j)})}> \frac{u_i(s_i, \sigma_{-i})-u_i(s'_i, \sigma_{-i})}{u_j(s_j, \sigma_{-j})-u_j(s'_j, \sigma_{-j})}$, we have
\[\delta_i(\Gamma)\,\big(u_j(s_j, \sigma_{-j})-u_j(s'_j, \sigma_{-j})\big)(\mathds{1}\{s_i\in F_i(\Gamma)\}-\mathds{1}\{s'_i\in F_i(\Gamma)\})\]
\[> \delta_j(\Gamma)\,\big(u_i(s_i, \sigma_{-i})-u_i(s'_i, \sigma_{-i})\big)(\mathds{1}\{s_j\in F_j(\Gamma)\}-\mathds{1}\{s'_j\in F_j(\Gamma)\}),\]
which implies that either $\mathds{1}\{s_i\in F_i(\Gamma)\}-\mathds{1}\{s'_i\in F_i(\Gamma)\}>0$ or $0>\mathds{1}\{s_j\in F_j(\Gamma)\}-\mathds{1}\{s'_j\in F_j(\Gamma)\}$.

\subsection{Proof of Observation \autoref{obs1}} \textbf{Part (i).} Since $s^*_i$ is a weakly dominant strategy, $u_i(s^*_i, \textbf{s}_{-i})\ge \max_{s'_i\in S_i\setminus s_i} u_i(s'_i, \textbf{s}_{-i})$. Therefore, $R_i(s^*_i)\le 0$. However, for any $s_i\in S_i\setminus{s^*_i}$, $u_i(s_i, \textbf{s}_{-i})\le u(s^*_i, \textbf{s}_{-i})=\max_{s'_i\in S_i\setminus s_i} u_i(s'_i, \textbf{s}_{-i})$. Therefore, $R_i(s_i)\ge 0$ for any $s_i\in S_i
\setminus{s^*_i}$. Therefore, $s^*_i\in F_i(\Gamma)$.

\medskip
\noindent\textbf{Part (ii).} If $s_i$ weakly dominates $s'_i$, then we have $R_i(s_i)\le R_i(s'_i)$. Since $s'_i\in F_i(\Gamma)$, equivalently, $R_i(s'_i)\le \overline{R}_i(\Gamma)$, we have $R_i(s_i)\le \overline{R}_i(\Gamma)$, i.e., $s_i\in F_i(\Gamma)$.

\subsection{Proof of Observation \autoref{obs2}} 

\noindent\textbf{Part 1.} Notice that $R_i(\bar{s}_i)=\max_{\textbf{s}_{-i}\in S_{-i}}\big(\max_{s'_i\in S_i\setminus \bar{s}_i} u_i(s'_i, \textbf{s}_{-i})-u_i(\bar{s}_i, \textbf{s}_{-i})\big)\le \max_{s\in S\setminus \bar{\textbf{s}}} u_i(s)-u_i(\underline{\textbf{s}})$. Moreover, for any $s_i\neq \bar{s}_i$, 
\begin{align*} R_i(s_i)=\max_{\textbf{s}_{-i}\in S_{-i}}\big(\max_{s'_i\in S_i\setminus s_i} u_i(s'_i, \textbf{s}_{-i})-u_i(s_i, \textbf{s}_{-i})\big) &\ge \max_{\textbf{s}_{-i}\in S_{-i}}\big(u_i(\bar{s}_i, \textbf{s}_{-i})-u_i(s_i, \textbf{s}_{-i})\big) \\ 
\ge u_i(\bar{s}_i, \bar{\textbf{s}}_{-i})-u_i(s_i, \bar{\textbf{s}}_{-i}) &\ge u_i(\bar{\textbf{s}})-\max_{s\in S\setminus{\bar{\textbf{s}}}}u_i(s).\end{align*} 
Finally, since $u_i(\bar{\textbf{s}})>2\max_{s\in S\setminus \bar{\textbf{s}}} u_i(s)-u_i(\underline{\textbf{s}})$, we have 
\[R_i(\bar{s}_i)\le \max_{s\in S\setminus \bar{\textbf{s}}} u_i(s)-u_i(\underline{\textbf{s}})\le u_i(\bar{\textbf{s}})-\max_{s\in S\setminus{\bar{\textbf{s}}}}u_i(s)\le R_i(s_i).\]
Therefore, $\bar{s}_i\in F_i(\Gamma)$.\medskip

\noindent\textbf{Part 2.} Notice that $R_i(\underline{s}_i)=\max_{\textbf{s}_{-i}\in S_{-i}}\big(\max_{s'_i\in S_i\setminus \underline{s}_i} u_i(s'_i, \textbf{s}_{-i})-u_i(\underline{s}_i, \textbf{s}_{-i})\big)\ge \max_{s'_i\in S_i\setminus s_i} u_i(s'_i, \underline{\textbf{s}}_{-i})-u_i(\underline{s}_i, \underline{\textbf{s}}_{-i})\ge \min_{s\in S\setminus \bar{\textbf{s}}} u_i(s)-u_i(\underline{\textbf{s}})$. Moreover, for any $s_i\neq \bar{s}_i$, $R_i(s_i)=\max_{\textbf{s}_{-i}\in S_{-i}}\big(\max_{s'_i\in S_i\setminus s_i} u_i(s'_i, \textbf{s}_{-i})-u_i(s_i, \textbf{s}_{-i})\big)\le u_i(\bar{\textbf{s}})-\min_{s\in S\setminus \bar{\textbf{s}}} u_i(s)$. Finally, since $u_i(\underline{\textbf{s}})< 2\min_{s\in S\setminus \bar{\textbf{s}}} u_i(s)-u_i(\bar{\textbf{s}})$, we have 
\[R_i(\underline{s}_i)\ge \min_{s\in S\setminus \bar{\textbf{s}}} u_i(s)-u_i(\underline{\textbf{s}})>u_i(\bar{\textbf{s}})-\min_{s\in S\setminus \bar{\textbf{s}}} u_i(s)\ge R_i(s_i).\]
Therefore, $\underline{s}_i\not\in F_i(\Gamma)$.

\bibliographystyle{ecta}
\bibliography{econref}

\end{document}